\documentclass[twocolumn,showpacs,preprintnumbers,amsmath,amssymb,superscriptaddress]{revtex4}

\usepackage{graphicx}
\usepackage{dcolumn}
\usepackage{bm}

\begin{document}

\preprint{Phys. Rev. E \textbf{78}, 041129 (2008).}

\title{Effect of memory on the prisoner's dilemma game in a square lattice}

\author{Shao-Meng Qin}
\affiliation{Institute of Theoretical Physics, Lanzhou University,
Lanzhou $730000$, China}

\author{Yong Chen}
\altaffiliation{Corresponding author. Email: ychen@lzu.edu.cn}
\affiliation{Institute of Theoretical Physics, Lanzhou University,
Lanzhou $730000$, China} \affiliation{Key Laboratory for Magnetism
and Magnetic materials of the Ministry of Education, Lanzhou
University, Lanzhou $730000$, China}

\author{Xiao-Ying Zhao}
\affiliation{Institute of Theoretical Physics, Lanzhou University,
Lanzhou $730000$, China}

\author{Jian Shi}
\affiliation{Institute of Theoretical Physics, Lanzhou University,
Lanzhou $730000$, China}

\date{\today}

\begin{abstract}
We have studied the effect of memory on evolution of the prisoner's
dilemma game using square lattice networks. Based on extensive
simulations, we found that the density of cooperators was enhanced
by an increasing memory effect for most parameters. However, we also
observed that the density of cooperators decreased with an increased
memory effect in the case of a large memory and moderate temptation.
It is interesting to note that memory makes cooperators immune from
temptation. The strength of protection reaches its maximal value
only for the moderate memory effect.
\end{abstract}

%02.50.Le Decision theory and game theory
%05.50.+q Lattice theory and statistics
%87.23.Ge Dynamics of social systems
%64.60.Ht Dynamic critical phenomena

\maketitle

\section{\label{introduction}introduction}

The evolutionary prisoner's dilemma game (PDG) has attracted
substantial attention over the past few decades~\cite{Axelrod}. In
this game, two agents must simultaneously select one of two
strategies: cooperation or defection. The prisoners receive payoffs
that are dependent on their choices. A selfish agent will adapt
their strategy to maximize their payoff. Game theory involves the
construction of many types of models and analysis of these models
using varied parameters. Therefore, game theory serves as a powerful
metaphore for simulation of the interactions between individuals in
many domains, including biology, economy, and ecology.

In the PDG, mutual cooperation generates the highest return for the
community. However, the Nash equilibrium state is mutual defection
because defection is a better choice for the prisoner, regardless of
the strategy of the other prisoner. Importantly, in the real world,
mutual cooperation is the most commonly utilized strategy. Systems
such as the PDG are considered to be an important tool for study the
emergence of cooperative behavior between selfish
individuals~\cite{2,3,4}. Nowak and May~\cite{Nowak} introduced a
spatial prisoner's dilemma game (SPDG) consisting of a two-state
cellular automaton. In the general SPDG, the agents in the game play
the PDG with their network neighbors and get payoffs according to a
payoff matrix. The total payoff of each agent is the sum of all
payoffs in this step. An agent may then mimic his neighbor's
strategy by comparing his payoffs in this step with his neighbor's
payoffs. An important conclusion is that spatial structure can
promote the persistence of cooperation. Because the interactions of
an agent are limited to his local neighbors, PDG models have been
extensively explored in the past few
years~\cite{Axelrod,book1,Net1,Net2}. In addition to spatial
structure, there are several mechanisms that may facilitate the
emergence and persistence of cooperation among populations. Hamilton
found that kin selection can favor cooperation~\cite{kin}. Axelrod's
model demonstrated that the~\textit{tit for tat} strategy could
sustain cooperation in systems of all players playing the game
together. The simulation performed by Szab\'o, Vukov, and Szolnoki
provided evidence that noise and irrational choices affect the
maintenance of cooperative behavior~\cite{noise}.

In the traditional SPDG model, the changing probability of strategy
is determined by the agents' performance on one step. In other
words, people assume that the agents are shortsighted and forgetful.
In fact, when people make an important decision, they generally
consider the current situation and their experiences. Therefore, the
effect of memory should be taken into account. Historical memory
plays a key role in the evolutionary game~\cite{WBH}. The purpose of
this paper is to evaluate whether memory enhances the density of
cooperators and detracts the cooperators from the temptation. We
observed the maximum value of critical points from a homogeneous
cooperator to a mixed state of cooperator and defector.

In this paper, we consider an evolutionary SPDG with the memory
effect in a square lattice, in which players update their strategy
by considering previous payoffs. The rules of the game are explained
in section II. The simulations, which are detailed in section III,
show that the evolution of SPDG depends on the magnitude of the
memory effect and payoff-matrix elements. Conclusions are drawn in
the last section.

\section{\label{model}Model}

In the traditional PDG, there are two players. Each player choses
one of two strategies: \textit{cooperator} ($C$) or
\textit{defector} ($D$). There are four combinations for the two
players: ($C$, $C$), ($C$, $D$), ($D$, $C$), and ($D$, $D$), which
corresponded to payoffs ($R$, $R$), ($S$, $T$), ($T$, $S$), and
($P$, $P$). The rewards or punishments for each player can be
tabulated as $2\times2$ payoff matrices (see Table \ref{table-1}).

\begin{table}
\caption{The payoff matrix of prisoner's dilemma game.}
\begin{ruledtabular}
\begin{tabular}{c|c c }
 player1$\setminus$player2 & \textit{C}   &\textit{D} \\ \hline
  \textit{C} & $ R \setminus R$ & $S \setminus T$ \\
  \textit{D} & $ T \setminus S$ & $P \setminus P$ \\
\end{tabular}
\end{ruledtabular}
\label{table-1}
\end{table}

Four elements in the payoff matrix satisfy the order ranking
$T>R>P>S$ and the additional constraint $T+S<2R$ for repeated
interactions. As suggested by Nowak and May~\cite{Nowak}, the
parameters in this paper are $R=1$, $T=b$, $S=0$, and $P=0$. Our
model preserve the essentials of PDG and $b$ is the only tunable
parameter.

Our study is based on systematic Monte Carlo (MC) simulations on a
square lattice network with periodic boundary conditions. When we
applied the PDG on the network, the players were located on the
nodes. In every MC step, the players simultaneously play the PDG
with their network neighbors (only the first neighborhoods) and
themselves. The sum payoff of each player is the sum over all games.
The evolutionary process is governed by strategy imitation. In every
MC step, all agents may mimic their neighbors strategy. Player $i$
adopts a (randomly chosen) neighbor's strategy (at site $j$) with a
probability that depends upon the payoff difference:

\begin{equation}
W = \frac{1} {1 + \exp \left[ (E_m(i)-E_m(j))/\kappa \right]},
\label{eq1}
\end{equation}

where $\kappa$ indicates the noise generated by the players allowing
irrational choices~\cite{ka1,PRE5869}. In this work, we use
$\kappa=0.1$ for all simulations. $E_m(i)$ and $E_m(j)$ are the
total payoffs which contain the sum payoffs at this MC step $U$ and
the cumulative historical payoff. For each node $i$, there are two
memories $M_c(i,t)$ and $M_d(i,t)$ at step $t$. When node $i$ is
associated with the strategy $C$ and the sum payoff at this MC step
is $U$,
\begin{eqnarray}
E_m(i) &=& U+M_c(i,t) \nonumber \\
M_c(i,t+1) &=& (M_c(i,t)+U)*\tau \nonumber \\
M_d(t+1) &=& M_d(t)*\tau, \label{eq-2}
\end{eqnarray}
for this time step. When the node $i$ is associated with strategy
$D$
\begin{eqnarray}
E_m(i) &=& U+M_d(i,t) \nonumber \\
M_c(i,t+1) &=& M_c(i,t)*\tau \nonumber \\
M_d(i,t+1) &=& (U+M_d(i,t))*\tau. \label{eq-3}
\end{eqnarray}

Here, $\tau$ is the memory factor and $0 \leq \tau < 1$. $M_c(i,t)$
and $M_d(i,t)$ represent the historical payoffs of $C$ and $D$,
respectively. The memory effect for each MC step declines with time.
In other words, memories of the payoffs, $M_c(i,t)$ and $M_d(i,t)$,
will be forgotten as time passes. $\tau =0$ indicates that there is
no memory effect. As $\tau$ nears to $1$, there exists an almost
perfect memory effect in the model. Starting from a random initial
state with an equal fraction of $C$ and $D$ and
$M_c(i,0)=M_d(i,0)=0$, we iterate the model with a synchronized
update.

\section{\label{sec2}Simulation Results}

Our simulations are carried out by varying $b$ and $\tau$. The
results described in this paper are obtained from MC simulations
with a system size of $200 \times 200$, with the exception of the
results shown in Fig.~\ref{figbc}. It is true that a network with
larger size will decrease the ensemble error, which is caused by the
finite scale of networks. We have simulated our model with $100
\times 100$ and $400 \times 400$. There is not conspicuous
difference among these networks. The results in this manuscript are
the average of $20$ trials with various random seeds. Repeating
simulations with different random seeds can also reduce the error.
Therefore, the $200 \times 200$ is large enough. The transient time
is varied from $20,000$ to $80,000$ MC steps. After the transient
state, the system reached the stable state, and the amplitudes of
population fluctuations were considerably smaller than the
corresponding average value.

\begin{figure}
\begin{center}
\includegraphics[width=0.5\textwidth]{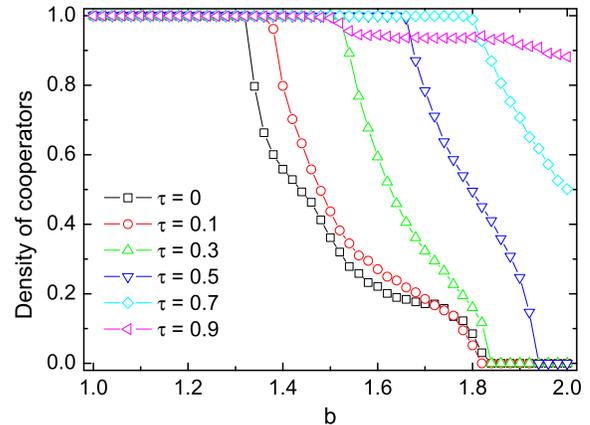}
\caption{\label{fig1}(Color online) Density of cooperators $f_c$ as
a function of the payoff parameter $b$ with various memory factors
$\tau$.}
\end{center}
\end{figure}

\begin{figure}
\begin{center}
\includegraphics[width=0.5\textwidth]{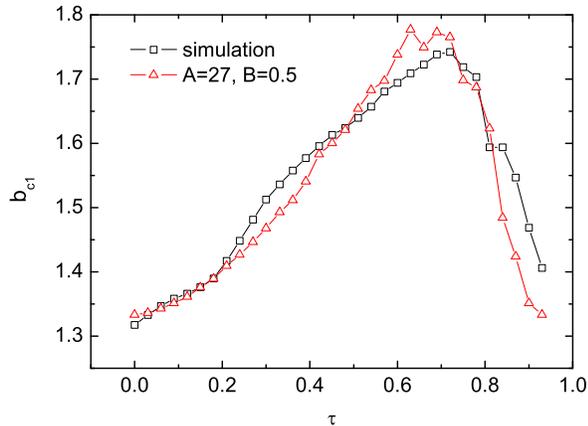}
\caption{\label{fig2}(Color online) $b_{c1}$ as a function of memory
factor $\tau$ on the square lattice. The data points depicted by
squares (black) are the result of MC simulations, and the data
points depicted by triangles (red) were derived from Eq.
(\ref{eq-5}). }
\end{center}
\end{figure}

To characterize the macroscopic behavior of the system, we measure
the density of $C$ $f_c$ first. Fig.~\ref{fig1} shows the $f_c$ on
the square lattice as a function of $b$ for several values of
$\tau$. We find that there are two thresholds of temptation $b$.
When $b<b_{c1}$, the networks in which only $C$ can survive are in
the stable state. The density of $C$ decreases monotonously with
increasing values of $b$ for $b>b_{c1}$. We upload the movies which
presents how the systems with network size of $240 \times 240$
evolve in 300 MC steps after transient time on different $b$ and
$\tau$ ~\cite{movie}. In these movies $C$s are presented by black
boxes and $D$s are presented by red boxes. It is observed that the
agents who utilize the same strategies join together to form complex
patterns that continuously move and change shape. These patterns
develop because agents change their strategies by learning from
their neighbors. Furthermore, the $C$s who join together are more
stable because they support each other by earning payoff from their
$C$ neighbors. For $b>b_{c2}$, $C$ strategies die out. Both the
memory factors $\tau$ and $\kappa$ affect the critical
point~\cite{noise}. Recently in Ref.~\cite{exponent2}  Szab\'o,
Vukov and Szolnoki draw a $\kappa-b$ plane of Newman-Watts networks.
In contrast to $\tau$, $\kappa$ does not conspicuously affect
$b_{c1}$ or $b_{c2}$ in this model. The main focus of this paper is
to evaluate how the memory effect $\tau$ affects the density of $C$
and $b_{c1}$. Determination of $\kappa-b_{c1}$ and $\kappa-b_{c2}$
is beyond the scope of this paper. $b_{c2}$ increases with $\tau$
monotonously; however, $b_{c1}$ reaches its maximum value near $\tau
=0.72$, and $b_{c1}$ tends toward $4/3$ as $\tau$ approaches $0$ or
$1$ (see black squares in Fig.~\ref{fig2}). Fig.~\ref{fig1}, we find
the memory effect enhances the density of $C$ in most cases;
however, the density of $C$ decreases with increasing of $\tau$ only
for $\tau > 0.72$ and $1.75 < b < 1.8$. It should be noted that our
simulations are consistent with those presented in Fig. 1 of
Ref.~\cite{PRE5869} for $\tau=0$ despite the fact that Szab\'o and
T\"oke used the asynchronized update law in their model. The
mean-field results for six-point approximations~\cite{5813} agree
with the simulation in~\cite{PRE5869} and our model in the case of
$\tau=0$. We assume that the six-point approximation includes the
main features of the two models. Importantly, the six-point
approximation does not contain a restriction of the update law.
Therefore, it is conjectured that the synchronized update does not
play an important role in the two models.

\begin{figure}
\begin{center}
\includegraphics[width=0.5\textwidth]{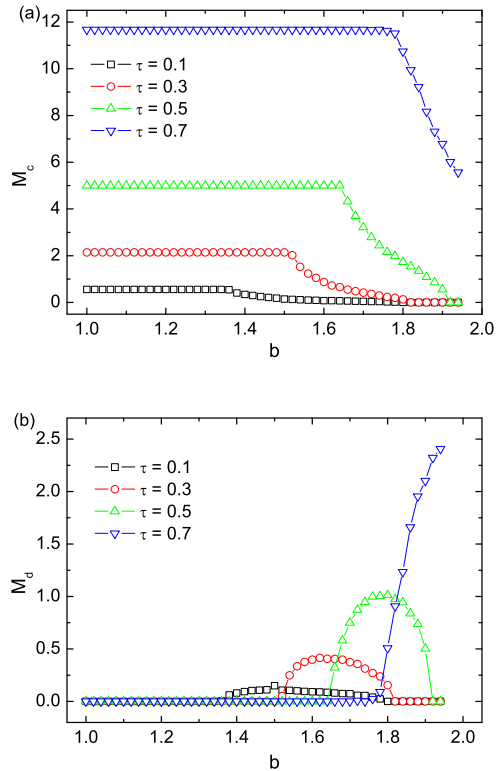}
\caption{\label{fig3}(Color online) The average payoffs for
strategies $C$ and $D$ as a function of the payoff parameter $b$ for
several values of the memory factor $\tau$.}
\end{center}
\end{figure}

In comparision to the case of $\tau =0$, we know that enhancement of
the density of $C$ is caused by $M_c$ and $M_d$. From the
above-mentioned definitions, the $M_c$ and $M_d$ of one node are
determined by two factors: (1) the payoff income $U$ of every MC
step and (2) whether the node maintains one strategy. $M_c$ or $M_d$
is aggravated if the node persists in $C$ or $D$ respectively.
Fig.~\ref{fig3} plots the average $M_c$ and $M_d$ of all nodes as a
function of $b$. It should be noted that $M_c$ is always larger than
$M_d$. Therefore memory effect almost always enhances $f_c$ in this
model. For $b<b_{c1}$, the networks include only $C$. Every node can
receive payoffs $5$ at every MC step and $M_c$ is $5*\tau/(1-\tau)$.
Then, with an increase of $b$, the emergence of $D$ reduces the
value of $C$'s payoff for every MC step and decreases the continuous
accumulation of $M_c$. As a result, $M_c$ gradually decreases with
$b$ until $C$ dies out and $M_c=0$. In contrast to $M_c$, $M_d$ has
a peak in the $C$-$D$ coexistent states. When $D$ is outside of the
mixed region, $M_d$ is equal to $0$. $D$ earns a payoff only by
playing the game with $C$. Therefore, $M_d$ is not equal to 0 in the
$C$-$D$ coexistent region $b_{c1}<b<b_{c2}$. When $b$ is little bit
larger than $b_{c1}$ and $1-f_c \ll 1$, $D$ forms small isolated
\textit{gangs}. As discussed in~\cite{PRE5869}, the behaviors of $D$
\textit{gangs} are considered as branching and annihilating random
walkers~\cite{18,randomwalk}. The $D$ \textit{gangs} undergo four
basic processes: random walk; an annihilation reaction (two $D$
\textit{gangs} can unite); death (one \textit{gang} of $D$ will die
due to the irrational choice); and branching (one \textit{gang} of
$D$ can divided into two gangs). Every $D$ gang that obtains the
highest payoff at every MC step is surrounded by cooperators.
However, the density of $D$ is low, and random walking breaks the
continual accumulation of $M_d$. Therefore, $M_d$ is small. When $D$
is dominant, the random walking of $C$ \textit{gangs} does not
deplete the accumulating of $M_d$ but the average payoff of $D$
decreases at each MC step. Thus $M_d$ is maximized when there is a
compromise between the average payoff at each MC step and continual
accumulation of $M_d$.

\begin{figure}
\begin{center}
\includegraphics[width=0.5\textwidth]{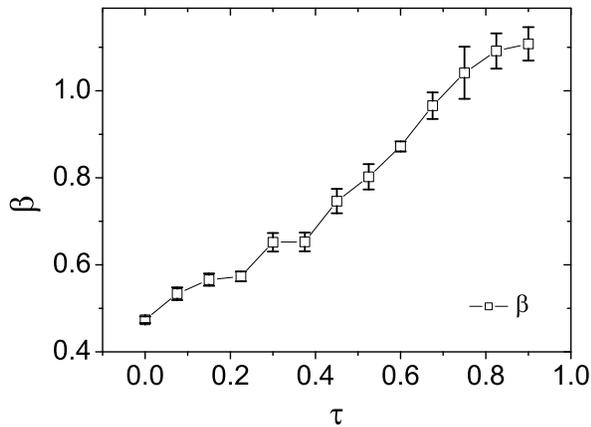}
\caption{\label{figbc}The critical exponent $\beta$ as a function of
$\tau$ of $b_{c1}$. The error bar in the figure presents the
standard deviation. In order to suppress the statistical error in
the critical regions, we use the system size $600\times600$ for
$\tau\leq0.6$, $800\times800$ for $0.75\geq\tau>0.6$ and
$1000\times1000$ for $\tau >0.75$.}
\end{center}
\end{figure}

In ~\cite{PRE5869,exponent2}, the authors discussed the critical
exponent of $b_{c1}$ and $b_{c2}$. Their MC simulations indicated a
power-law behavior, namely $f_c \propto (b_{c2}-b)^{\beta}$ and
$1-f_c \propto (b-b_{c1})^{\beta}$, and the values of $\beta$ agreed
with the directed percolation (DP) exponent. Grassberger and Janssen
conjectured that all one-component models with a single absorbing
state belong to the universality class of (DP) ~\cite{58_19}. The
value of critical exponents should be independent of the details of
dynamical rules and dependent on the spatial dimension. In this
paper, we investigated these exponents in the context of different
values of $\tau$. Fig.~\ref{figbc} shows that $\beta$, which ranged
from 0.47 ($\tau=0$) to 1.10825 ($\tau=0.9$) is monotonously
increase with $\tau$. Therefore, the value of the critical exponent
is not universal but depends on the memory factor $\tau$ in this
model.

\begin{figure}
\begin{center}
\includegraphics[width=0.5\textwidth]{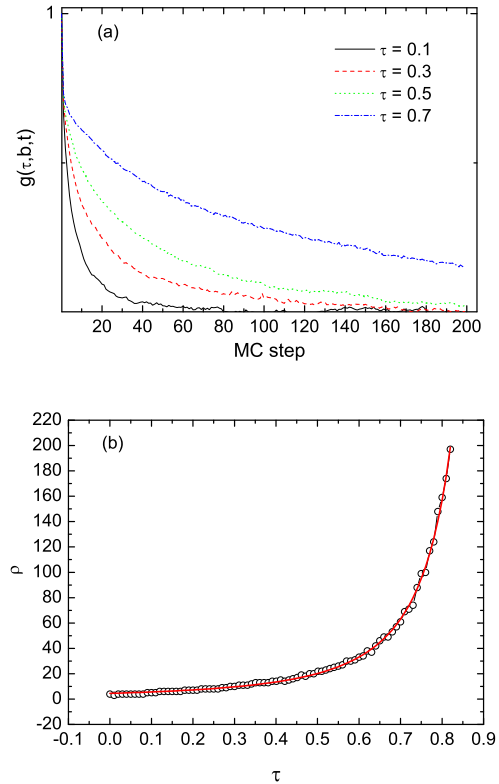}
\caption{\label{fig5} (color online) (a) Time autocorrelation
function of strategy for several values of $\tau$. (b) The
characteristic time $\rho(\tau)$ as a function of $\tau$. The red
line is the fitting result of this figure
$\rho=4.29*(1-\tau)^{-2.22}$}
\end{center}
\end{figure}

Considering that a persistent unchanged strategy at one site leads
to the accumulation history payoff, we investigated the mobility of
spatial patterns of $\tau$. Population mobility is a central feature
of real ecosystems: animals migrate, bacteria run and tumble.
Similar phenomena can be observed in a rock-paper-scissors game
~\cite{natureMobility}. Reichenbach, Mobilia, and Frey, observed
that mobility critical influence on species diversity. In this
model, we find that the behavior of $b_{c1}$ is caused by the
decrease in strategy mobility. This means that $C$ resists
temptation $b$ by decreasing mobility. Therefore, we introduce the
time autocorrelation function of strategy:

\begin{equation}
g(\tau,b,t)= \left< s_i(0)s_i(t) \right>, \label{eq-4}
\end{equation}

where $s_i(t)$ is the strategy of player $i$ at MC step $t$. When
player $i$ chose $C$, $s_i(t)=1$. In contrast, $s_i(t)=-1$ for $D$.
$\left< \right>$ denotes an average over all nodes in the networks.
Considering that $g(\tau,b,t)$ can be affected by density of $C$ and
in order to ensure that $g(\tau, b, t)$ ranges from $0$ to $1$, we
chose $b$ such that $f_c=0.5$. This definition describes whether the
node's recent strategy correlates with its strategy at later $t$ MC
steps.

Fig.~\ref{fig5}(a) displays the attenuation of $g(\tau,t)_{f_c=0.5}$
with time. It was found that $g(\tau,t)_{f_c=0.5}$ fits with the
form $g(\tau,t)_{f_c=0.5} = \exp (-t/\rho(\tau))$. One can regard
$\rho$ as the characteristic residence time of the unaltered
strategy. We define $t_h$ as the number of MC steps for which one
strategy was maintained and assume that the characteristic residence
time $\rho$ and $t_h$ have a similar ratio:

\begin{equation}
t_h=\rho/A+B, \label{eq-bc}
\end{equation}

Fig.~\ref{fig5}(b) shows $\rho$ as a function of the parameter
$\tau$. There is a critical behavior at $\rho \propto
(1-\tau)^{-z}$, where the exponent is $z=2.22$ with standard
deviation $0.043$.

\begin{figure}
\begin{center}
\includegraphics[width=0.5\textwidth]{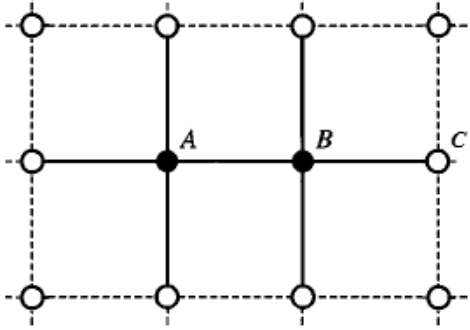}
\caption{\label{fig6} Illustration of the $D$-$D$ pair (nodes A and
B) and neighbor $C$ (node C). The black circle and white circle
denote $D$ and $C$, respectively. $D$-$D$ pair indicates that both
nodes are connected in the networks by strategy $D$.}
\end{center}
\end{figure}

Now we focus our attention on the behavior of $b_{c1}$. When
$b>b_{c1}$, $C$ cannot resist temptation $b$ and $D$ appears.
Therefore, $b_{c1}$ can be regarded as the ability of the model to
protect $C$. As described in the discussion above, the $D$
\textit{gangs} undergo four basic processes. When $b=b_{c1}$ and
$1-f_c \ll 1$, the annihilation process is rare, while the death and
branching processes are major activities. Therefore, $D$
\textit{gangs} become stable if the branching rate is greater than
the death rate. We found that the single $D$ in the branching
process will have an offspring and form $D$-$D$ pairs (as shown in
Fig.~\ref{fig6}). The $D$-$D$ pair plays an important role in the
branching process of $D$ \textit{gangs}. When we discard the effect
of noise, the total payoff $E_m$ of each player in $D$-$D$ pairs
(nodes A and B in Fig.~\ref{fig6}) must be larger than the payoff of
his neighbor $C$ (node C in Fig.~\ref{fig6}). Otherwise, the $D$
\textit{gangs} will eventually die. For example, in the case of
$\tau =0$, the total payoff of each player in $D$-$D$ pairs is $3b$,
and the total payoff of his neighbor $C$ is $4$. Therefore, the
threshold for a stable $D$ is $b_{c1}=4/(3b)=4/3$. We suggest that
the deviation of $4/3$ , which was obtained our simulations was
caused by noise.

The behavior of a stable $D$ is subtle for $\tau>0$. Based on the
discussion above, the increase of $t_h$ with $\tau$ and $t_h$
determines the player's memory and total payoffs. Therefore, we can
use $t_h$ to approximate $b_{c1}$. When the $t_h$ of the $D$-$D$
pair is $N$ and we neglect the remnants $M_d$, which accumulated
many MC steps ago and assume that neighbor $C$ can remain as $C$
indefinitely because of the dominance of $C$ at $b=b_{c1}$, we find
that

\begin{equation}
b_{c1}=\frac{4}{3(1-t_h^{N+1})}. \label{eq-5}
\end{equation}

In Fig.~\ref{fig2}, we plot the results from Eq.~(\ref{eq-5}) which
are similar to the simulation results. We use $A=27$ and $B=0.63$ in
Eq.~(\ref{eq-bc}).

\section{\label{sec3}Conclusion}

In this paper, we studied the ability of memory to protect $C$ from
$D$ in an evolutionary PDG in a square lattice networks. With an
increase in the effect of memory, there is an increase in the
density of $C$ in most cases. In compution of the autocorrelation
function, we used the characteristic residence time to measure the
mobility of a spatial pattern. We also found that the mobility of a
spacial pattern decreases with the memory effect. Decreasing
mobility induces a maximum value of critical coexistence point
$b_{c1}$ at $\tau=0.72$. It is obvious that mobility plays an
important role in this model. The effect of memory on cooperative
behaviors may draw some attention in evolutionary games.

We have also applied this model to the Newman-Watts small-world
(NWSW) networks~\cite{NWSW}. The NWSW network is a two dimension
small-world network. We found that moderate long range links did not
have an obvious qualitative influence on our model.

\begin{acknowledgments}
X.Y.Z. and J.S. acknowledge financial support of the National Talent
Training Fund in Basic Research. Y.C. was supported by the National
Natural Science Foundation of China under Grant No. $10305005$ and
by the Fundamental Research Fund for Physics and Mathematic of
Lanzhou University. This study was supported by HPC program of kd-50
in University of Science and Technology of China.

\end{acknowledgments}

\end{document}